\documentclass[a4paper]{jpconf} 
\usepackage{graphicx}
\usepackage{epsfig}
\usepackage{bm}
\usepackage{amsfonts}

\usepackage{color}

\newcommand{\fix}[2]{#1}

\newbox\pippobox

\def\be{\begin{equation}}
\def\ee{\end{equation}}
\def\ba{\begin{eqnarray} }
\def\ea{\end{eqnarray}}

\newcommand {\lla} {\ {\raise-.5ex\hbox{$\buildrel<\over\sim$}}\ }
\renewcommand{\(}{\left(}
\renewcommand{\)}{\right)}

\usepackage[T1]{fontenc}
\usepackage[latin1]{inputenc}
\usepackage{graphicx}
\usepackage[english]{babel}
\usepackage{amsmath}
\usepackage{amssymb}
\usepackage{amsfonts}

\begin{document}

\title{Generalized Three-Form Field}

\author{Pitayuth Wongjun$^{1,2}$}
\address{$^1$ The institute for fundamental study, Naresuan University, Phitsanulok 65000, Thailand}
\address{$^2$ Thailand Center of Excellence in Physics, Ministry of Education, Bangkok 10400, Thailand}
\ead{pitayuthw@nu.ac.th}

\begin{abstract}
A generalized three-form \fix{field}{filed} is an extended version of the canonical three-form field by considering a \fix{}{generic} Lagrangian of the \fix{generalized}{} three-form field as a function of \fix{the}{} kinetic and \fix{the}{} mass terms. In this work, we investigated cosmological models due to this generalized three-form field. It is found that one can use the three-form field to interpret the non-relativistic matter without \fix{the}{} caustic problem. Moreover, by analyzing the dynamical system, a viable model of dark energy due to the generalized three-form field can be obtained.
\end{abstract}




\section{Introduction}
It is well-known that there are two accelerated expansion phases of the Universe: the inflationary phase, taking place during the very early era of the Universe, and the dark energy phase, corresponding to the expansion nowadays. A scalar field model is \fix{one of}{} simple model\fix{s}{} to describe these phenomena \cite{ArmendarizPicon:2000dh,ArmendarizPicon:2000ah,Chiba:1999ka,Boubekeur:2008kn}. Beside the cosmological models due to the scalar field, a three-form field can be successfully used to describe both inflationary \fix{phase}{models} and dark energy \fix{phase}{models} \cite{Germani:2009iq,Koivisto:2009sd, Kobayashi:2009hj,Germani:2009gg,Koivisto:2009fb, Koivisto:2009ew,Ngampitipan:2011se,DeFelice:2012jt, Koivisto:2012xm, Kumar:2014oka, Barros:2015evi, DeFelice:2012wy, Koivisto:2011rm, Wongjun:2016tva, Kumar:2016tdn, Morais:2016bev}. At the perturbation level, it is obvious to see that the three-form field can generate intrinsic vector perturbations while it is not possible for the scalar field. In the present work, by mimicking the k-essence scalar field, we investigate a generalized version of the three-form field by considering a \fix{}{generic} Lagrangian as a function of kinetic and mass terms \cite{Wongjun:2016tva}. We briefly review the important ingredients of the three-form field including energy momentum tensor and the equation of motion in covariant form. The cosmological solution of the model is investigated. A constant equation of state parameter yields the power law of both kinetic and mass terms in the Lagrangian so that one can use the three-form field to interpret the non-relativistic matter without caustic problem which is a problem found in k-essence scalar field \fix{model}{}. This issue is discussed in section \ref{sec:model} in the present work. In section \ref{sec:dynamics}, we consider the non-constant case of the equation of state parameter and then use the dynamical system to analyze the behavior of the Universe. We found that it is possible to obtain a viable model of dark energy due to the generalized three-form field. Finally, the results are summarized and discussed in section \ref{summary}.

\section{Generalized Three-Form Filed \label{sec:model}}
In this section, we will review the generalized three-form field model by considering a \fix{}{generic} Lagrangian of the three-form field,  $A_{\alpha\beta\gamma}$, as a function of the kinetic and mass terms as follows\fix{,}{} \cite{Wongjun:2016tva}
\begin{equation}
S=\int d^{4}x\sqrt{-g}\left[\frac{M_{Pl}^2}{2}R + P(K,y)\right],\label{action-th}
\end{equation}
where the kinetic term and scalar quantity of the three-form field are expressed as
\begin{eqnarray}
K &=&-\frac{1}{48}\,
F_{\alpha\beta\gamma\delta}F^{\alpha\beta\gamma\delta},\quad y = \frac{1}{12} A_{\alpha\beta\gamma}A^{\alpha\beta\gamma},\quad
F_{\mu\nu\rho\sigma}=\nabla_{[\mu}A_{\nu\rho\sigma]}.
\end{eqnarray}
From this action, the equations of motion and the energy momentum tensor of the three-form can be written as
\begin{eqnarray}
E_{\alpha\beta\gamma} &=&\nabla_\mu\left(P_{,K} F^\mu_{\,\,\,\alpha\beta\gamma}\right) + P_{,y}A_{\alpha\beta\gamma} = 0, \label{eom-co}\\
T_{\mu\nu} &=& \frac{1}{6}P_{,K}F_{\mu\rho\sigma\alpha}F_\nu^{\,\,\,\rho\sigma\alpha}-\frac{1}{2} P_{,y} A_{\mu\rho\sigma}A_{\nu}^{\,\,\,\rho\sigma}+P g_{\mu\nu}. \label{EM-co}
\end{eqnarray}
where the notation with subscript $P_{,x}$ denotes  $P_{,x} = \partial_x P$.
The energy momentum tensor is conserved up to the equation of motion as $6 \nabla_\mu T^{\mu}_{\,\,\nu}= F_{\nu\alpha\beta\gamma}E^{\alpha\beta\gamma} = 0.$

By considering a flat Friedmann-Lemaitre-Robertson-Walker (FLRW) metric, the components of the three-form field, $A_{\alpha\beta\gamma}$, can be written as $A_{0ij}=0, A_{ijk} =a^3\varepsilon_{ijk} \, X(t)$ where $\varepsilon_{ijk}$ is the three-dimensional Levi-Civita symbol with $\varepsilon_{123}=1$. As a result, the energy density and pressure of the three-form \fix{field}{} can be expressed as
\begin{eqnarray}
\rho_X &=& 2K P_{,K} -P,\label{rho}\\
p_X &=& -\rho_X - 2y P_{,y}. \label{pressure}
\end{eqnarray}
For the equation of state parameter, $w_X = p_X /\rho_X =-1-2 y P_{,y}/\rho_X$, a partial differential equation of $P$ can be written  as
\begin{eqnarray}
2 y P_{,y} + (1+w_X) 2K P_{,K} = (1+w_X) P. \label{eq-Pform}
\end{eqnarray}
By taking the equation of state parameter to be a constant, one can solve this equation to obtained the exact solution as $P=P_0 K^{\nu} y^{\mu}$, where $P_0$ is an integration constant and  $\mu$, $\nu$ are the exponent constants. As a result\fix{,}{} the equation of state parameter can be expressed in terms of the constants as
\begin{eqnarray}
w_X = -1 + \frac{2\mu}{1-2\nu}, \,\,\,\, \nu \neq \frac{1}{2}.
\end{eqnarray}
From this expression, the description of non-relativistic matter can be obtained by choosing a proper parameters to yield $w_X=0$, for example, $\mu = 1, \nu = -1/2$. Therefore, the Lagrangian of the three-form is still finite and then the model can avoid the caustic problem which is found in the k-essence scalar field case \cite{Wongjun:2016tva}. Moreover, one can relax the assumption of constant $w$ by assuming $w_X = w_X(y)$. As a result, one of the solutions can be written as $P=P_0 K^{\nu} e^{\frac{(1-2\nu)}{2}\lambda y}$ \cite{Wongjun:2016tva} where $w_X= -1 + \lambda y$ and $\lambda$ is a constant. From this solution, we found that it is possible to obtain the solution which admits the dark energy model. We will investigate this possibility in detail by using the dynamical system in \fix{the}{} next section.

\section{Dynamical system of the dark energy model\label{sec:dynamics}}
\begin{figure}[h!]
\begin{center}
\includegraphics[scale=0.45]{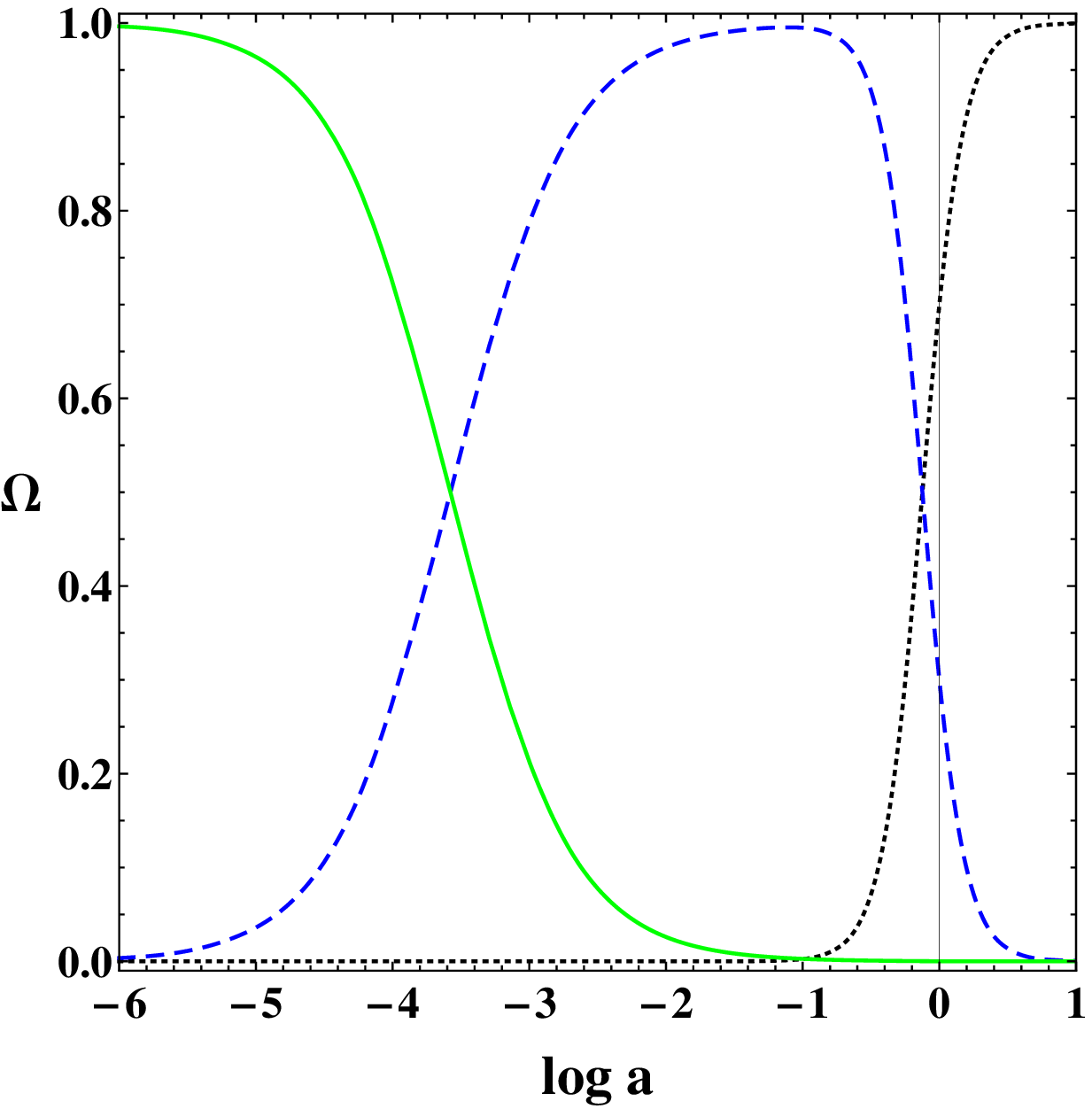}\qquad
\includegraphics[scale=0.45]{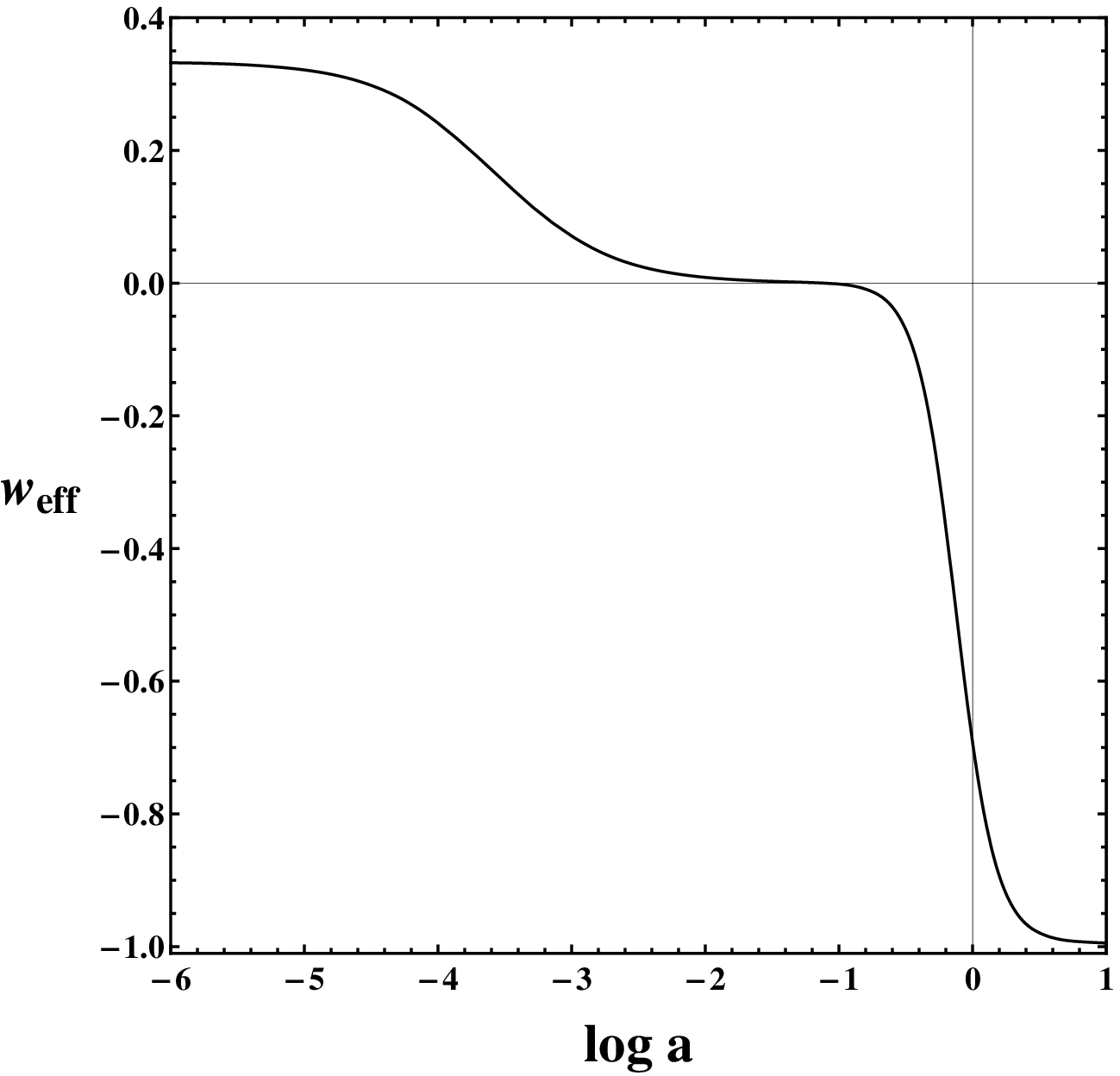}
\end{center}
{\caption{The left panel shows the evolution of the density parameters for three-form dark energy (\fix{b}{B}lack dotted line), matter (\fix{b}{B}lue dashed line) and radiation (\fix{g}{G}reen solid line). The right panel shows the evolution of the effective equation of state parameter $w_{eff}$. The parameters are set as $\alpha = -0.1$ and $n=1$. }\label{fig:evo}}
\end{figure}

In this section, the dynamics of the Universe will be explored by using the dynamical system. By including the non-relativistic matter and radiation into the generalized three-form field model in Eq. (\ref{action-th}), non-zero components of Einstein equation can be written as
\begin{eqnarray}
1 &=& \Omega_X + \Omega_m + \Omega_r,\label{EinEq1}\\
\frac{2 H'}{3H^2} &=& -1-w_X \Omega_X -\frac{1}{3} \Omega_r,\label{EinEq2}
\end{eqnarray}
where $\Omega$ is the density parameter, $H = \dot{a}/a$ is the Hubble parameter, prime denotes the derivative with respect to $N = \ln a$ and the subscripts $m$ and $r$ represent the quantity corresponding to matter and radiation respectively. The autonomous system is obtained by using the conservation \fix{of}{} energy momentum tensor for each species together with the Einstein equations, Eq. (\ref{EinEq1}) and Eq. (\ref{EinEq2}). As a result, the autonomous system corresponding to the generalized three-form field can be written as
\begin{eqnarray}
\Omega_X' &=&  -3\Omega_X \(w_X(1-\Omega_X) - \frac{1}{3}(1-\Omega_X - \Omega_m)\),\\
\Omega_m' &=& 3\Omega_m \(w_X\Omega_X + \frac{1}{3}(1-\Omega_X - \Omega_m)\),\\
w_X'&=& (1+w_X)\Gamma \frac{y'}{y},\,\, \Gamma = 1-\frac{y Q_{,y}}{Q}+\frac{y Q_{,yy}}{Q_{,y}},\label{evo-wx}
\end{eqnarray}
where we have chosen $P= P_0 K^\nu Q(y)$. From this system, one can see that the equation of $w_X$ is decoupled from the others. Note that $\Gamma = 0$ for $Q\propto y^\mu$ which \fix{implies}{in turn} that $w'_X = 0$ as we expect for the case of constant $w_X$. From the autonomous system, there are three fixed points as follows; matter-dominated fixed point $(\Omega_X,\Omega_m, \Omega_r) = (0,1,0)$, radiation-dominated fixed point $(\Omega_X,\Omega_m, \Omega_r) = (0,0,1)$ and dark-energy-dominated fixed point $(\Omega_X,\Omega_m, \Omega_r) = (1,0,0)$. Considering  equation of $w_X$ in Eq. (\ref{evo-wx}), it is convenient to choose the solution such that $y'/y = \alpha$ where $\alpha$ is a constant. This choosing is qualitatively similar to one in usual three-form field model \cite{Koivisto:2009ew,Ngampitipan:2011se,DeFelice:2012jt}. Therefore, the equation for $w_X$ in Eq. (\ref{evo-wx}) has a fixed point at $w_X = -1$ and its dynamics and stability depend on \fix{the}{} sign of $\alpha \Gamma$. If $\alpha \Gamma < 0$, \fix{where}{the} $w_X =-1$\fix{,}{and} the dark-energy-dominated fixed point\fix{}{s} will be \fix{a}{} stable fixed point\fix{}{s}.  By choosing $Q$ \fix{to be}{as} a more general function than one in \cite{Wongjun:2016tva} as $Q = \exp\(\frac{(1-2\nu)}{2}\lambda y^n\)$, we obtain $\Gamma = n$.  The conditions to avoid ghost and Laplacian instabilities have been investigated in  \cite{Wongjun:2016tva} and then provide the conditions to the model parameter\fix{}{s} as $n(2\nu - 1) >0$ and $n(2+\lambda y^n) > 1$. The sufficient conditions for healthy \fix{}{of the} theory are $n \geq 1/2, \nu > 1/2, \lambda > 0$. Hence, $\Gamma$ \fix{must be}{is} always positive to avoid the instabilities. In order to obtain a viable model of dark energy, $\alpha$ must be negative. Moreover, the value of $|\alpha|$ must be sufficiently small to guarantee that $w_X$ \fix{does}{is} not deviate too much over the history of the Universe. We numerically show this viable model by setting $\alpha = -0.1$ and $n=1$ as found in Fig \ref{fig:evo}. Note that for $n \geq 2$, the result is not significantly different as long as $|\alpha|$ is sufficiently small enough.

\section{Summary\label{summary}}
In this paper, we investigated the generalized three-form field by considering a \fix{more}{most} general Lagrangian as $L = P (K,y)$ \cite{Wongjun:2016tva} where $K$ is a canonical kinetic term and $y$ is a standard mass term of the three-form field. From the energy momentum tensor of the three-form \fix{field}{}, we found the relation \fix{between}{of} $P$ and its partial derivatives via the equation of state parameter, $w_X$. By assuming constant $w_X$, the equation can be exactly solved and it solution can provide the \fix{}{a} description of non-relativistic matter without caustic problem as found in k-essence scalar field. For non-constant $w_X$, we specialize our attention in the case $w_X = w_X(y)$ and a suitable Lagrangian is chosen as $P= P_0 K^\nu \exp\(\frac{(1-2\nu)}{2}\lambda y^n\)$. By analyzing dynamical system, we found that it is possible to obtain a viable model of dark energy due to the generalized three-form field. We also confirm our analysis by using numerical method which can be found in Fig. \ref{fig:evo}.

Even though the model can be used to explain the cosmic acceleration of the Universe, it cannot provide the way to solve the coincidence problem. It is of interest to investigate dark energy and dark matter coupling to alleviate the coincidence problem in which both contents are constructed from the three-form fields. We leave this investigation for further work.

\ack{
The author is supported by Thailand Research Fund (TRF) through grant TRG5780046. The author would like to thank Lunchakorn Tannukij for value discussion and comments.
}


\section*{References}

\begin{thebibliography}{99}

\bibitem{ArmendarizPicon:2000dh}
  Armendariz-Picon~C, Mukhanov~V~F and Steinhardt~P~J
 2000 
 A Dynamical solution to the problem of a small cosmological constant and late time cosmic acceleration
  {\it Phys.\ Rev.\ Lett.\ }  {\bf 85} 4438 


\bibitem{ArmendarizPicon:2000ah}
  Armendariz-Picon~C, Mukhanov~V~F and Steinhardt~P~J
  2001
  Essentials of k essence
  {\it Phys.\ Rev.\ } D {\bf 63} 103510 

\bibitem{Chiba:1999ka}
  Chiba~T, Okabe~T and Yamaguchi~M
  2000
  Kinetically driven quintessence
  {\it Phys.\ Rev.\ } D {\bf 62} 023511 

\bibitem{Boubekeur:2008kn}
  Boubekeur~L, Creminelli~P, Norena~J and Vernizzi~F
  2008
  Action approach to cosmological perturbations: the 2nd order metric in matter dominance
  {\it JCAP} {\bf 0808} 028 




\bibitem{Germani:2009iq}
  Germani~C and Kehagias~A
  2009
  P-nflation: generating cosmic Inflation with p-forms
  {\it JCAP }{\bf 0903}, 028 

\bibitem{Koivisto:2009sd}
  Koivisto~T~S, Mota~D~F and Pitrou~C
  2009
  Inflation from N-Forms and its stability
  {\it JHEP } {\bf 0909}, 092 (2009)


\bibitem{Kobayashi:2009hj}
  Kobayashi~T and Yokoyama~S
  2009
  Gravitational waves from p-form inflation
  {\it JCAP } {\bf 0905}, 004 

\bibitem{Germani:2009gg}
  Germani~C and Kehagias~A
  2009
  Scalar perturbations in p-nflation: the 3-form case
  {\it JCAP } {\bf 0911}, 005 

\bibitem{Koivisto:2009fb}
  Koivisto~T~S and Nunes~N~J
  2009
  Inflation and dark energy from three-forms
  {\it Phys.\ Rev.\ }D {\bf 80}, 103509 

\bibitem{Koivisto:2009ew}
  Koivisto~T~S and Nunes~N~J
  2010
  Three-form cosmology
  {\it Phys.\ Lett.\ }B {\bf 685}, 105 (2010)

\bibitem{Ngampitipan:2011se}
  Ngampitipan~T and Wongjun~P
  2011
  Dynamics of three-form dark energy with dark matter couplings
  {\it JCAP }{\bf 1111}, 036 


\bibitem{DeFelice:2012jt}
  De Felice~A, Karwan~K and Wongjun~P
  2012
  Stability of the 3-form field during inflation
  {\it Phys.\ Rev.\ }D {\bf 85}, 123545 


\bibitem{Koivisto:2012xm}
  Koivisto~T~S and Nunes~N~J
  2013
  Coupled three-form dark energy
  {\it Phys.\ Rev.\ }D {\bf 88}, 123512 

\bibitem{Kumar:2014oka}
  Kumar~K~S, Marto~J, Nunes~N~J and Moniz~P~V
  2014
  Inflation in a two 3-form fields scenario
  {\it JCAP } {\bf 1406}, 064 

\bibitem{Barros:2015evi}
  Barros~B~J and Nunes~N~J
  2016
  Three-form inflation in type II Randall-Sundrum
  {\it Phys.\ Rev.\ }D {\bf 93}, no. 4, 043512 

\bibitem{DeFelice:2012wy}
  De Felice~A, Karwan~K and Wongjun~P
  2012
  Reheating in 3-form inflation
  {\it Phys.\ Rev.\ } D {\bf 86}, 103526 

\bibitem{Koivisto:2011rm}
  Koivisto~T~S and Urban~F~R
  2012
  Three-magnetic fields
  {\it Phys.\ Rev.\ } D {\bf 85}, 083508 

\bibitem{Wongjun:2016tva}
  Wongjun~P
  2016
  A Perfect Fluid in Lagrangian Formulation due to Generalized Three-Form Field
  ({\it Preprint} arXiv:1602.00682 [gr-qc])

\bibitem{Kumar:2016tdn}
  Sravan Kumar~K, Mulryne~D~J, Nunes~N~J, Marto~J and Vargas Moniz~P
  2016
  Non-Gaussianity in multiple three-form field inflation
  {\it Phys.\ Rev.\ } D {\bf 94}, no. 10, 103504 

\bibitem{Morais:2016bev}
  Morais~J, Bouhmadi-Lopez~M, Sravan Kumar~K, Marto~J and Tavakoli~Y 
  2017
  Interacting 3-form dark energy models: distinguishing interactions and avoiding the Little Sibling of the Big Rip
  {\it Phys.\ Dark Univ.\ } {\bf 15} 7





 \end{thebibliography}
\end{document}